\documentclass[conference,10pt]{IEEEtran}
\usepackage{cite}
\usepackage{amsmath,amssymb,amsfonts,bm}
\usepackage{algorithmic}
\usepackage{graphicx}
\usepackage{textcomp}
\usepackage{xcolor}
\def\BibTeX{{\rm B\kern-.05em{\sc i\kern-.025em b}\kern-.08em
		T\kern-.1667em\lower.7ex\hbox{E}\kern-.125emX}}
\usepackage[letterpaper, portrait, margin=0.7in, bmargin=0.95in]{geometry}

\usepackage{fancyhdr}
\fancypagestyle{firststyle}{\fancyhf{}
	\fancyhead[L]{Dipankar Shakya, Ting Wu, and Theodore S. Rappaport ``A Wideband Sliding Correlator based Channel Sounder in 65 nm CMOS: An Evaluation Board Design",\textit{ in GLOBECOM 2020 - 2020 IEEE Global Communications Conference, }Taipei, Taiwan, Dec. 2020, pp. 1--6}
}

\begin{document}
	
\title{A Wideband Sliding Correlator based Channel Sounder in 65 nm CMOS: An Evaluation Board Design\\
}

\author{\IEEEauthorblockN{Dipankar Shakya}
	\IEEEauthorblockA{\textit{NYU WIRELESS} \\
		\textit{New York University}\\
		New York City, USA \\
		ds5981@nyu.edu}
	\and
	\IEEEauthorblockN{Ting Wu}
	\IEEEauthorblockA{\textit{Center for Neural Science} \\
		\textit{New York University}\\
		New York City, USA \\
		ting.wu@nyu.edu}
	\and
	\IEEEauthorblockN{Theodore S. Rappaport}
	\IEEEauthorblockA{\textit{NYU WIRELESS} \\
		\textit{New York University}\\
		New York City, USA \\
		tsr@nyu.edu}
}

\maketitle

\linespread{1.05}

\thispagestyle{firststyle}

\begin{abstract}
  Wide swaths of bandwidth at millimeter-wave (mm-Wave) and Terahertz (THz) frequencies stimulate diverse applications in wireless sensing, imaging, position location, cloud computing, and much more. These emerging applications motivate wireless communications hardware to operate with multi-Gigahertz (GHz) bandwidth, at nominal costs, minimal size, and power consumption. Channel sounding system implementations currently used to study and measure wireless channels utilize numerous commercially available components from multiple manufacturers that result in a complex and large assembly with many costly and fragile cable interconnections between the constituents and commonly achieve a system bandwidth under one GHz. This paper presents an evaluation board (EVB) design that features a sliding correlator based channel sounder with 2 GHz null-to-null RF bandwidth in a single monolithic integrated circuit (IC) fabricated in 65 nm CMOS technology. The EVB landscape provides necessary peripherals for signal interfacing, amplification, buffering, and enables integration into both the transmitter and receiver of a channel sounding system, thereby reducing complexity, size, and cost through integrated design. The channel sounder IC on the EVB is the world's first to report gigabit-per-second baseband operation using low-cost CMOS technology, allowing the global research community to now have an inexpensive and compact channel sounder system with nanosecond time resolution capability for the detection of multipath signals in a wireless channel.     
\end{abstract}


\section{Introduction}
Wireless communication technologies utilize mm-Wave frequencies in fifth-generation (5G) cellular wireless \cite{Rappaport2017itap,Dahlman2019spawc,Agiwal2016icst}, while future sixth-generation (6G) cellular wireless systems and beyond envision operation at even higher frequencies into the Terahertz (THz) range \cite{Rappaport2019ia}. The higher frequencies provide wide swaths of GHz range bandwidth that can be leveraged for diverse applications involving thousands of interconnected devices communicating at multi Gigabits-per-second (Gbps) data rates with sub-millisecond latency \cite{Akyildiz2014pc}. Especially in the study of wireless channels, GHz wide bandwidths allow systems to detect multipath signals at nanosecond time differences which facilitates development of accurate channel models for indoor and outdoor environments, and applications such as centimeter level accurate position location, environmental sensing, and imaging   \cite{xing2019indoor,kanhere2019map,xie2018mc,Rappaport2019ia}.  

The onward march into higher frequency ranges necessitates the improvement of existing RF hardware in terms of operating frequency, bandwidth, noise figure (degradation factor of signal-to-noise ratio by a device), sensitivity (minimum signal level detectable at the receiver), and monolithic integration of high performance components\cite{KO2019cm}. As an example, present foundry technologies are able to typically fabricate transceivers that operate up to 200 GHz with sub-GHz RF bandwidth, noise figures at 9 dB and power amplifiers that can provide 180 mW power\cite{KO2019cm,Rodwell2019spawc}. Evidently, the operational bandwidth of most commercial-off-the-shelf (COTS) wireless devices remains under one GHz and monolithic designs are pursuant of achieving higher bandwidth. Currently, most channel measurement setups also operate under one GHz bandwidth and involve the integration of multiple discrete COTS components including clock sources, amplifiers, mixers, signal generators, and transceivers \cite{Lee2015pimrc, Mac2017sac, Mac2017icc, Miao2015pimrc, Xing2018gc}. Alongside cost implications, complexity and interfacing requirements among multiple components from different manufacturers, COTS systems are limited by the maximum chip rate of the baseband electronics, time resolution in the power delay profile (PDP) of the signals received, and difficulty in implementing multiple transmitting and receiving channels in parallel for the study of multiple antenna wireless communications \cite{Miao2015pimrc, Lee2015pimrc, Mac2017icc, Wu2019iscas}.

\begin{figure*}[htbp]
	\centering
	\includegraphics[width=1\textwidth]{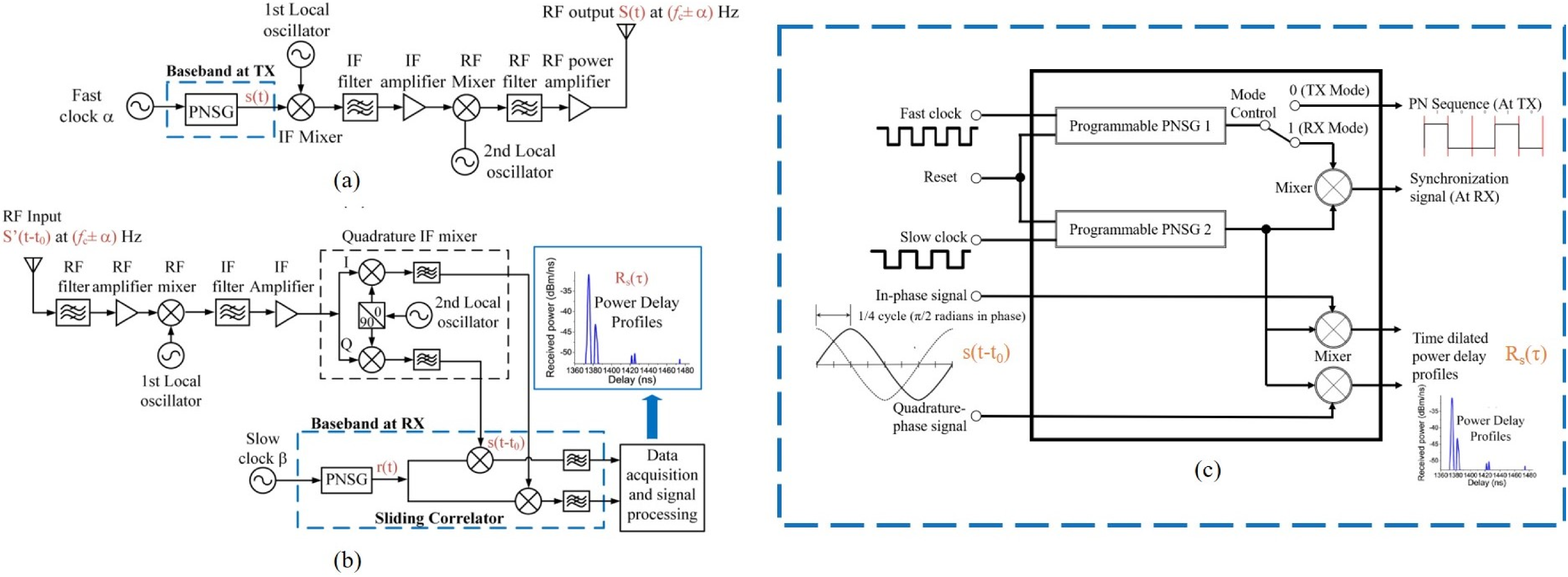}
	\caption{(a) A transmitter system block diagram that generates pseudo-random noise sequence at chip rate $\alpha$. (b) A receiver system block diagram with a sliding correlator implementation for generating time dilated power delay profiles. (c) Block diagram of the baseband at tansmitter and receiver merged on a single channel sounder chip \newline}
	\label{fig:Sl_corr}
	\vspace{-1.8em}
\end{figure*}
This paper presents an evaluation board (EVB) design that features an ultra-wideband, programmable, synchronization capable channel sounder integrated circuit (IC) implemented in 65 nm complementary metal oxide silicon (CMOS) technology as the principal component. The EVB introduces peripherals necessary to interface the IC into a channel sounding system including clock buffers, single ended and differential signal converters, amplifiers, programming switches and a power supply. The rest of the paper is organized as follows: Section \ref{sec2} describes fundamental concepts involved in designing the baseband monolithic IC, Section \ref{sec3} highlights the board design process, and Section \ref{Sec4} presents the final board fabrication with initial test results.           

\section{Design and specifications of the on-board sliding correlator based channel sounder chip}\label{sec2}

\subsection{Sliding correlator based channel sounding}
Introduced by Don Cox in \cite{Cox1972ap}, sliding correlation is a time tested method for the measurement of ultra-wideband wireless channels using direct sequence spread spectrum signals. The method involves transmission of a predetermined signal known both to the transmitter (TX) and receiver (RX) beforehand, with signal statistics that resemble Gaussian white noise, known as the pseudo-random noise (PN) sequence \cite{pirkl2008wc}. The RF signal received at RX is then correlated with the PN sequence replica which results in a power delay profile of the received signal.

Fig. \ref{fig:Sl_corr}(a) and (b) are block representations of TX and RX setup in a sliding correlator based channel sounder \cite{Wu2019iscas}. In reference to Fig. \ref{fig:Sl_corr}(a), initially a PN sequence generator (PNSG) generates a PN sequence, $s(t)$, at a chip rate equal to the fast clock input, $\alpha$. The PNSG is implemented using maximal linear feedback shift registers with programmable feedback taps. At the subsequent mixer, $s(t)$ modulates an intermediate frequency (IF) signal generated by the first local oscillator. The modulated signal at IF is upconverted to the RF carrier frequency, $f_{c}$, resulting in the RF output signal at $f_{c}\pm\alpha$ that is transmitted over the wireless channel. The wideband signal thus transmitted is subject to reflection, diffraction, scattering, and transmission through blockages in the channels before arriving at the receiver. The effects of the channel cause multiple replicas of the transmitted signal to arrive at the receiver over multiple paths at different delays. As illustrated in Fig. \ref{fig:Sl_corr}(b), the received multipath signals are down-converted and demodulated before sliding correlation is performed with a replica of the transmitted PN sequence operating at a slightly slower clock rate of $\beta$, referred to as $r(t)$. The result of the sliding correlation process is a time dilated PDP that represents the time domain response of the channel, given by: 
\begin{equation}\label{s_corr_op}
\begin{split}
R_{s}(\tau)&= \int_{0}^{T} r(t)s(t-t_{0}-\tau)dt,\\
\end{split}
\end{equation}

The dilation in the time scale of the received PDP corresponds to the sliding factor $\gamma$, obtained as in Eq. \eqref{gama}.              
\begin{equation}\label{gama}
\begin{split}
\gamma&= \frac{\alpha}{\alpha-\beta},\\
\end{split}
\end{equation}

\subsection{Design of the on-board channel sounder IC}
The dashed blue boxes in Fig. \ref{s_corr_op}(a) and (b) represent the baseband components of the TX and RX which is combined by the channel sounder chip design on a monolithic IC design that is capable of operating at either end of the channel sounder system. As shown in Fig. \ref{s_corr_op}(c), the `Mode Control' pin on the IC is used to switch between TX mode for the pin set to `Low' and RX mode for pin set to `High'.

The output of PNSG 1 directly generates PN sequence at chip rate of $\alpha$ in TX mode. In RX mode, the output of PNSG 1, $s(t)$, is mixed with the output of PNSG 2, $r(t)$ at chip rate $\beta$, to generate the synchronization signal which is a zero delay absolute timing reference for the multipath components arriving at the receiver. The signal $r(t)$ is also fed into two mixers in order to perform the sliding correlation function with the in-phase and quadrature components of the demodulated received signal. The output of the two mixers constitute the result of the sliding correlation function, $R_{s}(\tau)$.

\section{Design and Testing of the EVB components}\label{sec3}
\subsection{Channel Sounder IC Specifications based on Measurements}

\begin{figure}[htbp]
	\centerline{\includegraphics[width=0.45\textwidth]{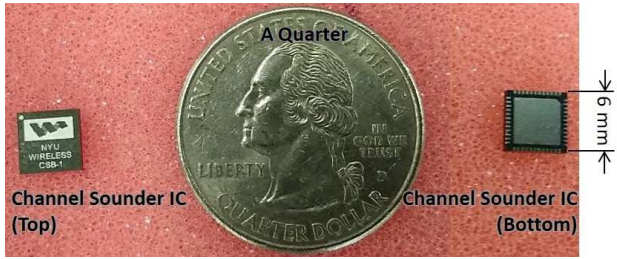}}
	\caption{The 1 mm $\times$ 0.66 mm channel sounder chip in 65 nm CMOS housed in a QFN48 IC package of 6 mm $\times$ 6 mm with 48 pin connections along the edges.}
	\label{fig:chips}
\end{figure}

Fabricated in a 65 nm CMOS process, the channel sounder chip is a monolithic design of 1 mm $\times$ 0.66 mm silicon, packaged in an industry standard QFN48 IC package of 6 mm $\times$ 6 mm, as displayed in Fig. \ref{fig:chips}, and is the centerpiece of the EVB design \cite{Wu2019iscas,post2006iedm,Lai2004tencon}. 

\begin{figure}[htbp]
	\centerline{\includegraphics[width=0.47\textwidth]{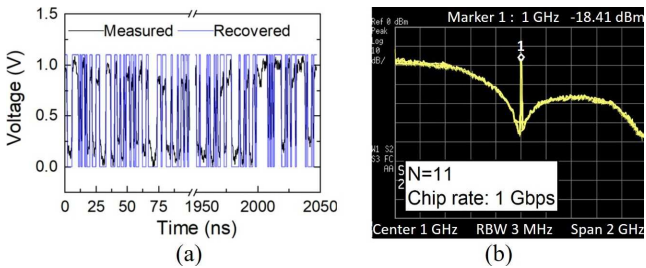}}
	\caption{(a)Time domain measurement of the generated PN sequence; analog waveform (black), bit sequence (blue). (b) Measured spectrum for the PN sequence operating at 1 Gbps for sequence length N=11}
	\label{fig:msmts}
	\vspace{-0.5em}
\end{figure}

Initial tests with the chip directly wire bonded to a printed circuit board (PCB), and a 1.1 V supply voltage verified the sequence accuracy, chip rate, and bandwidth for different lengths of PN codes. Upon feeding with a 1 GHz clock signal as $\alpha$ and programming a sequence length of 2047 ($2^{N}-1, N=11$), the chip sequence was observed in time domain as the black trace in Fig. \ref{fig:msmts}(a). Peak-transitions in the analog waveform higher than 0.5 V were recorded as `1', else `0' to obtain the blue trace and verify valid strings of same consecutive bits (runs) in the sequence. The spectrum measurement for the 1 Gbps PN sequence was as shown in Fig. \ref{fig:msmts}(b) with the power profile showing a distinct energy peak of -18.4 dBm at 1 GHz, indicating  2 GHz null-to-null RF bandwidth. Specifications of the chip following the initial testing are tabulated in Table \ref{tab:specs} \cite{Wu2019iscas}.  

\begin{table}[htbp]
	\caption{Measured specifications for the fabricated wideband channel sounder baseband design}
	\label{tab:specs}
	\begin{center}
		\begin{tabular}{|l|p{4cm}|}
			\hline
			\textbf{Item}&{\textbf{Specification}} \\
			\hline 
			\textbf{Technology} & 65 nm CMOS\\
			\hline
			\textbf{Maximum Chip Rate} & 1 Gbps\\
			\hline
			\textbf{Multipath Delay Resolution} & 1 ns\\
			\hline
			\textbf{Null-to-Null RF Bandwidth} & 2 GHz\\
			\hline
			\textbf{PN Code Length} & $2^{N}-1$ ($N=\{5,6,..,12\}$, \newline programmable)\\
			\hline
			\textbf{Power Consumption} & 6 mA at 1.1 V\\
			\hline
			\textbf{IC Packaging} & QFN48 6$\times$6 mm$^{2}$ \\
			\hline
			\textbf{Synchronization} & Supported\\
			\hline
		\end{tabular}
	\end{center}
	\vspace{-1.5em}
\end{table}

\subsection{Layout of the EVB}
\begin{figure}[htbp]
	\centerline{\includegraphics[width=0.45\textwidth]{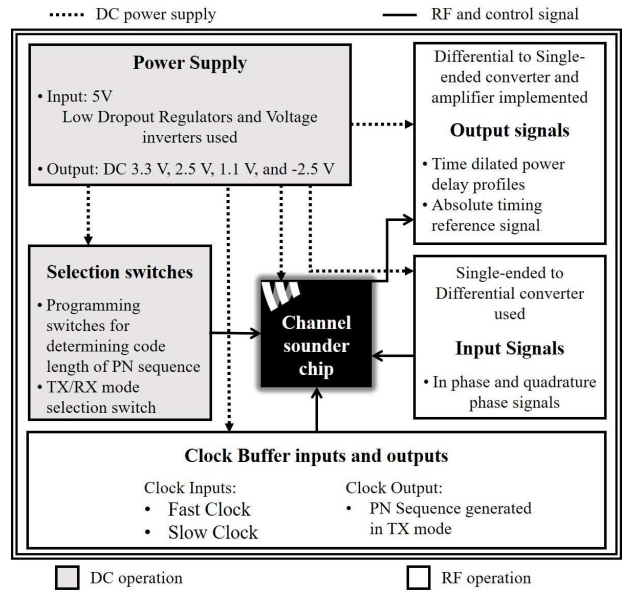}}
	\caption{Functional block diagram of the EVB sub-units with relative board positioning. Gray highlighted blocks represent DC operating frequency.}
	\label{fig:evb_layout}
\end{figure}

The EVB design focuses on interfacing the wideband channel sounder IC to the 142 GHz channel sounding setup at NYU WIRELESS \cite{Xing2018gc}. The board provides essential peripherals for the IC operation, which perform amplification and filtering of RF signals, inter-conversion between signal communication methods --differential and single-ended signaling, buffering and amplification of clock signals, and provide stable power supply at different voltages. Electronic components on the board are grouped into sub-units based on function and distributed over the board area according to the layout displayed in Fig. \ref{fig:evb_layout}.           

The board is laid out with sub-units operating at DC placed in a separate region from sub-units operating at frequencies up to 1 GHz. The separation of regions helps prevent interference on RF signal traces from the DC common-mode, and manage return paths within separated DC/RF ground planes. It helps preserve the integrity of signals near RX sensitivity (typically below -100 dBm) common at mm-Wave and THz frequencies.

\subsection{Design considerations for the EVB}
	\textbf{Grounding:} All signals flowing in a circuit seek the lowest impedance path to return to the signal power source to complete the circuit. As many signals are operating in a typical circuit, a single wire or trace of limited width and cross-section providing the return connection may have to carry hundreds of amperes current per $cm^{2}$ of cross-section area \cite{Wang2009ecce}. Thus, a ground plane is a standard method of providing a broad cross-section for returning signal currents while minimizing the return path resistance. However, it is not a panacea for providing return paths to all signals as the behavior of returning currents varies by frequency. A DC signal returns over the shortest path on the ground plane, while RF signals seek out the path of minimum impedance \cite{Zumbahlen2012ad}. For a two-layer PCB layout with signals routed on the top and ground on the bottom layer, RF signals above 1 MHz travel directly beneath the top layer signal trace. Furthermore, as DC signal return currents take the shortest path to ground, loops are formed on the ground plane, which may couple noise and interfere with RF signals.
	
	RF signal traces routed on multi-layer PCBs require a continuous ground plane underneath without any slits and breakages along the entire return path. The presence of disruptions in the ground plane changes the path length for RF signals, as returning signals loop around the breakage\cite{Pithadia2013aa1}. For example, at a millimeter wavelength, the change in path length by a half-mm may put differential signals in-phase and change the path impedance to cause complete signal reflections.                  
	\newline
	
	\vspace{-1.0em}
	\textbf{Impedance of RF signal traces:} Reflection of RF signals is avoided, and maximum power flow is maintained during the propagation of an RF signal over traces across the circuit board when the impedance is uniform throughout the signal path. Sub-miniature version-A (SMA) connectors with characteristic impedance of 50 $\Omega$ are commonly used to transport RF signals over the wire in RF systems. Thus, RF signal traces with a continuous ground underneath, called micro-strip traces, are designed to have 50 $\Omega$ characteristic impedance for RF-PCBs, as obtained from Eq. \eqref{im_eqn} \cite{graham1993ucb}.
	\begin{equation}
	\label{im_eqn}
	\begin{split}
		Z_{0} &= \frac{87}{\sqrt{\epsilon_{r}+1.41}} ln\left(\frac{5.98h}{0.8w+t}\right),\\
	\end{split}
	\end{equation}
	\vspace{-0.5em}	
	\begin{equation*}
	\begin{split}
		where,\\ Z_{0} =& \text{ micro-strip impedance} \\
		\epsilon_{r} =& \text{ dielectric constant of insulating material}\\
		h =& \text{separation between trace and ground in mils}\\ &\text{(1 mil= 0.001 inches)}\\
		w =& \text{width of the trace in mils}\\
		t =& \text{thickness of the trace in mils}\\
	\end{split}
	\end{equation*}

	The trace width ($w$) in Eq.\eqref{im_eqn} may be varied to achieve the 50 $\Omega$ impedance. An extension to Eq. \eqref{im_eqn}, shown in Eq. \eqref{diff_eqn}, is required to calculate the characteristic impedance for differential signal pairs as signals in a differential pair are referenced to each other and not to ground\cite{graham1993ucb}. Finally, the trace length determines the input impedance offered by the trace and the connected load.
	\begin{equation}
	\label{diff_eqn}
	\begin{split}
	Z_{d} &= \frac{174}{\sqrt{\epsilon_{r}+1.41}} ln\left(\frac{5.98h}{0.8w+t}\right)\left(1-0.48e^{\left(-0.96\frac{d}{h}\right)}\right),\\
	\end{split}
	\end{equation}
	\begin{equation*}
	\begin{split}
	where, \text{ $d$ is the separation of differential pair traces in mils} \\
	\end{split}
	\end{equation*}
	

\subsection{Sub-units of the EVB and Test Results}         
Before implementation on the final EVB, the sub-units were designed and tested on individual test PCBs. The test boards were fabricated in-house at NYU-MakerSpace on double-sided bare PCB with 18 $\mu$m copper layers each side separated by 0.79 mm flame retardant-4 (FR-4) insulation material with a dielectric constant of $\sim$4.2. The fabricated test boards are displayed in Fig. \ref{fig:tst_brd}. The sub-units are detailed as follows:

\begin{figure}[htbp]
 \centerline{\includegraphics[width=0.4\textwidth]{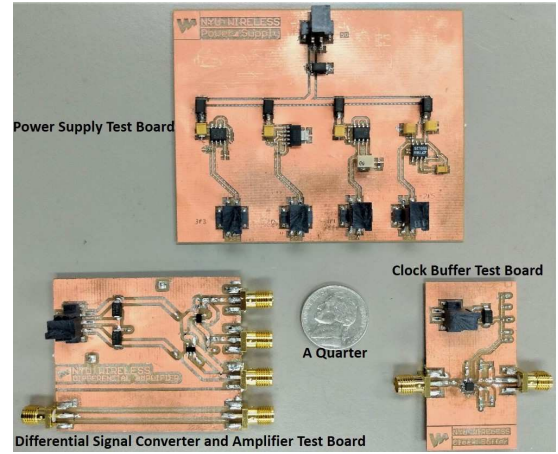}}
 \caption{In-house fabricated test boards on double sided PCB. RF connections terminated with SMA connectors, DC connections with wire terminal blocks.}
 \label{fig:tst_brd}
\end{figure}

	\textbf{Single-ended signal converter and amplifier:} To eliminate noise from the power supply, and thermal noise from external components that single-ended signals are subject to, the mixers and amplifiers in the chip baseband are designed for differential operation\cite{Razavi2002tm,Wu2019iscas}. However, the RF signals carried on coaxial cables interconnecting components of the 142 GHz channel sounder, setup at NYU WIRELESS are single-ended. Based on Fig. \ref{fig:Sl_corr}(b),  to interface the chip input with the preceding IF mixer at the RX, each single-ended input signal from the IF mixers are split into two signals of equal magnitude but opposite polarity using a 50 $\Omega$ 700 MHz balun circuit. The implemented balun is a transformer with a 1:1 turns ratio between the primary and center-tapped secondary winding\cite{jorgesen2014balun}. The primary is fed by the 500 MHz low pass filter output of the preceding IF mixers\cite{Xing2018gc}. The secondary winding terminals of the balun circuit provide the differential input signals required by the chip. 
	
	In initial tests, a 500 MHz 100 mVp-p square wave was split into out-of-phase 44 mVp-p sinusoids, indicating a 0.5 dB insertion loss.\newline
	
\begin{figure}[htbp]
		\centerline{\includegraphics[width=0.42\textwidth]{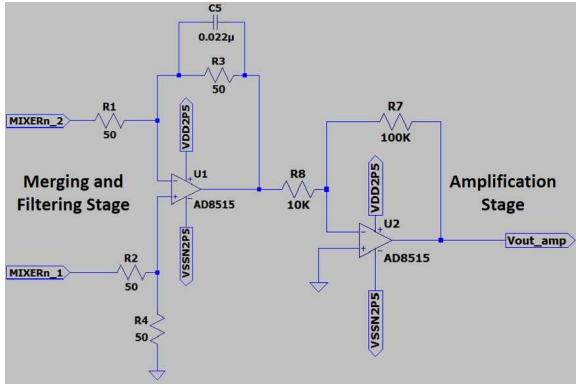}}
		\caption{Two stages of the differential to single ended signal converter and amplifier: the merging and filtering stage followed by the amplification stage; implemented with the AD8515 operational amplifier from Analog Devices.}
		\label{fig:diff_sch}
		\vspace{-1.5em}
	\end{figure}

	\vspace{-1.0em}
	\textbf{Differential signal converter and amplifier:} A sliding correlation on two PN sequences results in a low-frequency time dilated PDP component with distortion components at a frequency above $\alpha-\beta$. This waveform is output by the chip in two components, each as a differential signal pair as in Fig. \ref{fig:Sl_corr}(b). The differential signal converter-and-amplifier merges the output signal pair into a single-ended signal, then low-pass filters and amplifies the single-ended signal, using the schematic shown in Fig. \ref{fig:diff_sch}.
	
   The merging and filtering stage outputs a single-ended signal with an amplitude equal to the difference between the two input signal amplitudes. The resistor (R3) and capacitor (C5) placed in parallel on the feedback line in Fig. \ref{fig:diff_sch} implement a first-order low-pass filter with a 3 dB bandwidth of 150 kHz to suppress the distortion component of the waveform.  The amplification stage then amplifies the single-ended signal by gain equal to the ratio of resistances on the feedback line (R7) to the input line (R8) in Fig. \ref{fig:diff_sch}, thus implementing a gain of 10 dB. The final board design implements a switch-selectable gain of 3, 10, or 20 dB.\newline
    
    \vspace{-1.0em}
    \textbf{Clock Buffer:} Clock signals are crucial to the implementation of a sliding correlator based channel sounder as they determine the rate of the PN sequence, bandwidth of the transmitted signal, and time dilation factor corresponding to $\gamma$. Direct distribution of the clock from a clock-source into different board components may degrade the waveform of the clock signal due to the added loading, noise from the on-board electronics, and cabling losses. For instance, a 10 pF additional stray capacitance can add a 200 mVp-p ringing distortion on a 500 mVp-p square wave\cite{Zumbahlen2012ad}. Thus, the addition of a buffer between the clock source and clock dependent components can alleviate the waveform degradation and amplify the clock source prior supplying the components. 
    	
\begin{figure}[htbp]
    	\centerline{\includegraphics[width=0.42\textwidth]{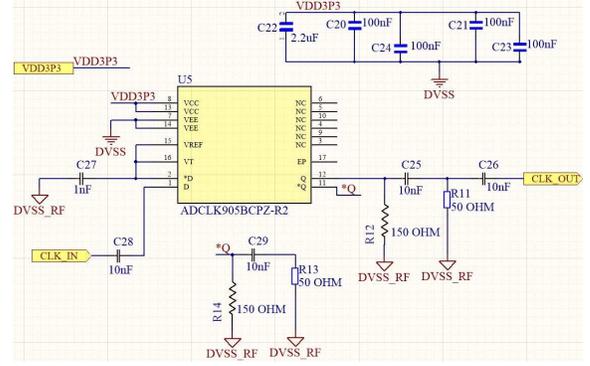}}
    	\caption{Clock buffer circuit schematic with matched terminations and DC blocking capacitor filters on clock input and output.}
    	\label{fig:clk_sch}
    	\vspace{-1.0em}
    \end{figure}

    A 7.5 GHz state-toggling capacity clock buffer is implemented as a test board based on Fig. \ref{fig:clk_sch}, for both the fast and slow clocks $\alpha$ and $\beta$ in Fig. \ref{fig:Sl_corr}(c). Testing showed a consistent 600 mVp-p square wave output at the input clock frequency for input down to 60 mVp-p.\newline
    
    \vspace{-1.0 em}
    \textbf{Power Supply:} The components of the EVB require DC supply voltage at different levels; the clock buffer operates at 3.3 V, the differential signal converter and amplifier operates at $\pm$ 2.5 V, and the wideband channel sounder IC requires 1.1 V for biasing transistors and 2.5 V for driving input and output pins on the IC. The three positive voltages are generated with low dropout (LDO) voltage regulators \cite{Fwu2018ti}, while the negative voltage is generated with a switched capacitor charge pump voltage inverter \cite{Bindra2012dk}. The current drawn by each of the test boards is measured with the power supply test board connected to a bench top regulated DC power supply at 5 V and the other boards connected to the voltage output ports of the power supply test board and is presented in Table \ref{tab:curr_draw}.              
	
	\begin{table}[htbp]
		\centering
		\caption{Current draw for each sub-unit of the EVB}
		\label{tab:curr_draw}
		\begin{tabular}{|p{2.8cm}|p{1cm}|p{1cm}|p{1cm}|p{1cm}|}
			\hline
			\textbf{Sub-Unit of EVB} & \textbf{Type} & \textbf{Supply voltage} & \textbf{Current draw (standby)} & \textbf{Current draw (active)}  \\
			\hline
			Power supply  & Active & 5 V & 4 mA & 172 mA \\
			\hline
			Clock buffer & Active & 3.3 V & $<$ 1 mA  & 44 mA\\
			\hline
			Differential signal converter and amplifier& Active & - 2.5 V/ 2.5 V & $<$ 1 mA & 5 mA \\
			\hline
			Single-ended signal converter & Passive & N/A & N/A & N/A\\
			\hline
			\multicolumn{5}{l}{N/A: Not applicable}
			
		\end{tabular}
	\vspace{-1.2em}
	\end{table}           
	           

\section{EVB Fabrication and initial measurements}\label{Sec4}
\begin{figure}[htbp]
	\centerline{\includegraphics[width=0.47\textwidth]{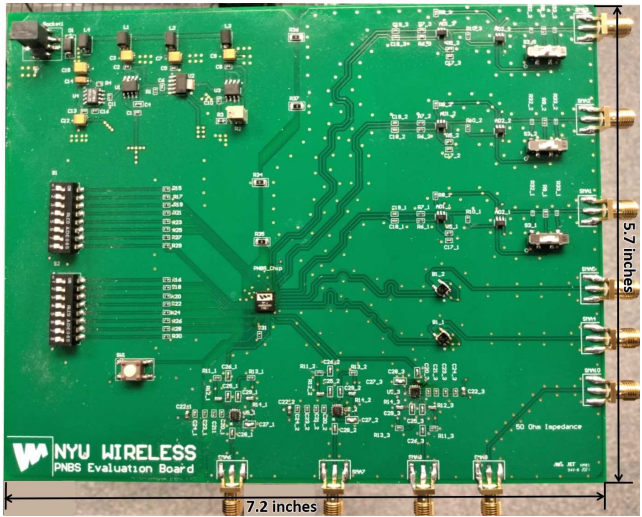}}
	\caption{Top view of the sliding correlator based channel sounder EVB with all components assembled. Channel sounder IC at the center.}
	\label{fig:fin_brd}
	\vspace{-1.0em}
\end{figure}

The final EVB is designed on a four-layer PCB with FR-4 dielectric insulation between layers. Micro-strip traces interconnecting components are routed on the top-layer with RF ground on the second layer plane. The third layer is used to distribute power supply lines and the bottom layer for grounding DC components in non-overlapping regions with the second layer. Further, the top-layer RF traces are shielded on either side by regions grounded with vias connected to the second layer RF ground. Such ground shielded traces minimize additional parasitic impedance from outward radiating fringing fields, prominent at mmWave and THz frequencies. Based on Eq. \eqref{im_eqn}, the PCB has a top-layer copper of thickness (t) 1.4 mils, a separation (h) of 9.13 mils between each layer with FR-4 insulation of $\epsilon_{r}$ 4.2. Hence, an impedance of 50.01 $\Omega$ is obtained for micro-strip traces with width (w) of 15.75 mils. Differential signal traces on the top-layer have a pair separation (d) of 25 mils for a differential impedance of 96.56 $\Omega$. The PCB features 50 $\Omega$ SMA connectors to interface with external components. The final PCB layout of the channel sounder EVB is presented in Fig. \ref{fig:fin_brd}. 

Following component assembly, EVB operation yielded a peak current draw of 172 mA at 5 V input when generating a transmit PN sequence at $\alpha$ of 1 GHz, with an energy peak of -30 dBm at 1 GHz in the measured spectrum power profile. The PN waveform showed consistent runs of ones and zeros, as described in \cite{Wu2019iscas}, at 600 mVp-p for $\alpha$ and $\beta$ input down to 70 mVp-p from 100 MHz to 1 GHz.

\section*{Conclusion}
A wideband sliding correlator based channel sounder IC fabricated using 65 nm CMOS processes is featured in an EVB design for interfacing with the 142 GHz channel sounder system at NYU WIRELESS. The monolithic channel sounder IC design, using low-cost CMOS technology, achieves a remarkable 1 Gbps baseband operation; the EVB for the IC implements the baseband of an entire channel sounder setup with null-to-null RF bandwidth of over one GHz. This design reduces channel sounder system costs from hundreds of thousands to a few thousand dollars while reducing system complexity, compressing size to portable volumes, and delivering higher resolution channel measurements. Future research involves interfacing the EVB and channel sounder system and measuring the board performance in operation.

\bibliographystyle{IEEEtran}
\bibliography{references}

\end{document}